\documentclass[twocolumn]{revtex4}
\usepackage{graphicx} 
\usepackage{color}
\usepackage{amsmath}
\begin{document}

\title{Atomic structures and electronic correlation of monolayer 1T-TaSe$_{2}$}

\author{Jae Whan Park$^1$}
\author{Han Woong Yeom$^{1,2}$}
\email{yeom@postech.ac.kr}
\affiliation{$^1$Center for Artificial Low Dimensional Electronic Systems,  Institute for Basic Science (IBS),  77 Cheongam-Ro, Pohang 790-784, Korea.} 
\affiliation{$^2$Department of Physics, Pohang University of Science and Technology, Pohang 790-784, Korea}
\date{\today}

\begin{abstract}

We investigate atomic and electronic structures of monolayer 1T-TaSe$_{2}$ using density functional theory calculations. 
Monolayers of 1T-TaSe$_2$ were recently grown on graphene substrates and suggested as an intriguing Mott insulator [Nat. Phys. {\bf 16}, 218 (2020)].
However, the prevailing structural model for the model system of 1T-TaS$_{2}$, the cation-centered cluster of a David-star shape with strong electron correlation, could not explain the characteristic and unusual orbital splitting observed in scanning tunneling spectroscopy experiments. 
We suggest an alternative structure model, an anion-centered cluster structure, which can reproduce most of the unusual spectroscopic characteristics with electron doping from the substrate without electron correlation. 
The unusual spectroscopic features observed, thus, seems to indicate a simple and usual band insulating state.
This work indicates the importance of a large structural degree of freedom given for a cluster Mott insulator. 

\end{abstract}


\maketitle


Mott insulator is a prototypical state of condensed matter with strong electron manybody interaction \cite{mott68}, which is related to a wide variety of exotic properties including unconventional superconductivity through the coupling to other types of manybody interactions \cite{cao18}. 
In particular, the commensurate charge-density-wave (CCDW) phase in 1T-TaS$_{2}$ \cite{brou80} is a paradigmatic example of a Mott insulator \cite{faze79} coupled with a strong electron-phonon interaction.
Among thirteen Ta atoms within a CCDW supercell called as a David-star cluster, twelve Ta atoms are bonded to each other but a unpaired 5$d$ electron in the central Ta atom falls into the Mott-insulator state due to a substantial onsite Coulomb repulsion and a very small band width.
This prevailing model is however under strong debate recently due mainly to the presence of interlayer coupling, which can induce an interlayer spin singlet and a trivial band insulating state.
In contrast, an isostructural and isoelectronic compound of 1T-TaSe$_{2}$ exhibits an equivalent CCDW structure to 1T-TaS$_{2}$ but remains in a metallic state down to very low temperature \cite{faze79, salv77,ge10}. 
This material has a greater hybridization of Ta 5$d$ and the Se 3$p$ electrons, which results in more diffuse Wannier orbitals for the unpaired electron, making it unfavorable to the Mott instability.
These are only parts of the studies which indicate the subtle balance between various different degrees of freedom to secure a Mott insulating state in a complex material. 

In this context, it is noteworthy that a very recent scanning tunneling spectroscopy (STS) study suggested a very unusual Mott insulating state in a monolayer of 1T-TaSe$_{2}$ \cite{chen20} grown on a graphene substrate through molecular beam epitaxy, which minimizes the interlayer coupling. 
The STS spectra of the monolayer 1T-TaSe$_{2}$ on a bilayer graphene/SiC(0001) substrate showed the $\sqrt{13}$$\times$$\sqrt{13}$ CDW superstructure shared by bulk 1T-TaSe$_{2}$ and 1T-TaS$_{2}$ and an apparent band gap.
This work assumed the David-star structure of bulk and local density of states (LDOS) calculated can reproduce the band gap with a moderate electron correlation (U $\sim$ 2 eV).
While this type of calculations in fact indicates a magnetically ordered state and the band gap is largely from the exchange interaction, this work interpreted it as the evidence of a Mott insulating state in line with many previous works on 1T-TaS$_{2}$. 
However, the putative upper Hubbard state shows a totally different orbital character from that expected for the unpaired $d_z$ electron on the central Ta atom.  
This observation substantially deviates from any Mott insulator known and an extra electronic interaction during electron tunneling was introduced to explain this unusual feature.  
In contrast, an independent STS study observed a different spectra for this state and suggested a band insulating state \cite{lin20}.


In this Letter, we note this discrepancy and the huge structural degree of freedom that a CDW unitcell with thirteen Ta atoms has.
In the latter aspect, we note further that the atomic structure of monolayer 1T-TaSe$_{2}$ films has not been determined experimentally. 
This means that the CDW unitcell can in principle be in one of various different structures competing energetically. 
The competition and coexistence of different CDW structures have been discussed for a long time for various CDW materials \cite{walk81,pore14,zhen18} and was recently observed, for example, in 2H-NbSe$_{2}$ \cite{gye19}.
Through DFT calculations, we suggest an atomic structure candidate of the CDW state of monolayer 1T-TaSe$_{2}$, the anion-centered structural model. 
This model can reproduce all the spectroscopic features observed including the unusual orbital texture of the unoccupied states with an electron doping from the substrate. 
The electron doping is believed to be highly possible in the electron-rich graphene bilayer grown on SiC. 
The electron correlation effect is shown to be marginal in the anion-centered model, indicating a band insulator.
This work tells that the picture of a Mott insulator with a unprecedented orbital texture has to be carefully examined and the atomic structure of the monolayer 1T-TaSe$_{2}$ film needs to be verified experimentally.

DFT calculations were performed by using the Vienna $ab$ $initio$ simulation package \cite{kres96} within the Perdew-Burke-Ernzerhof generalized gradient approximation \cite{perd96} and the projector augmented wave method \cite{bloc94}.
The monolayer 1T-TaSe$_{2}$ was modeled with the experimental lattice constant of 3.44 {\AA} \cite{chen20} and a vacuum spacing of about 13.6 {\AA}. 
We used a plane-wave basis set of 259 eV and a 8$\times$8$\times$1 $k$-point mesh for the  $\sqrt{13}\times\sqrt{13}$ unit cell and atoms were relaxed until the residual force components were within 0.01 eV/{\AA}.
To represent electronic correlations, an on-site Coulomb energy (U = 2.3 eV) was included for Ta 5$d$ orbitals.


\begin{figure}[t]
\centering{ \includegraphics[width=8.5 cm]{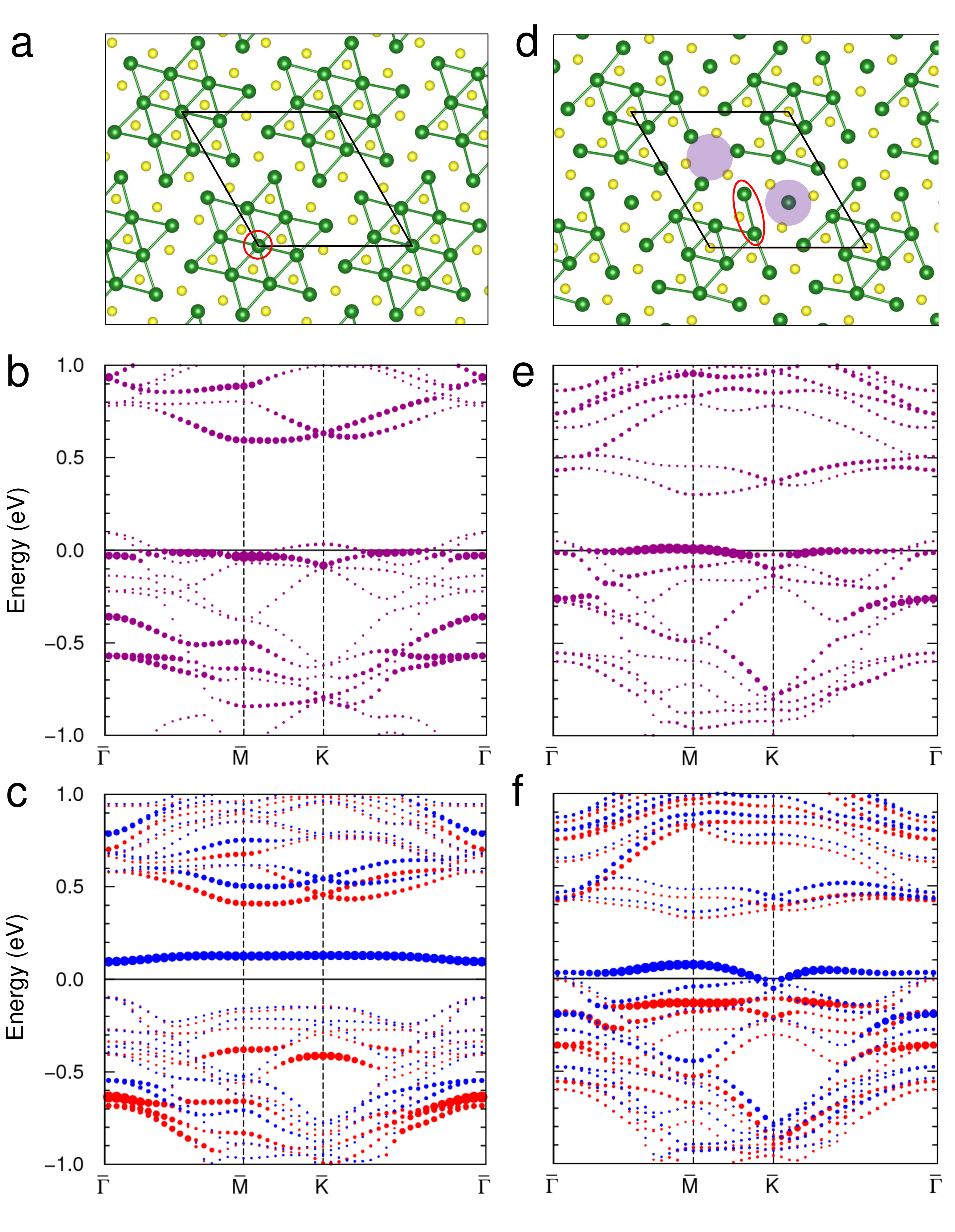} }
\caption{ \label{fig1}
Anion centered model of the monolayer 1T-TaSe$_{2}$.
{\bfseries\sffamily a-c} Cation centered model (David-star cluster) and {\bfseries\sffamily d-f} anion centered model.
{\bfseries\sffamily a} and {\bfseries\sffamily d} Atomic structures, {\bfseries\sffamily b} and {\bfseries\sffamily e} band structures without electron correlation, and {\bfseries\sffamily c} and {\bfseries\sffamily f} band structures with electron correlation (U = 2.3 eV).
Green and yellow balls in atomic structure represent Ta and top Se atoms, respectively.
The solid lines denote the $\sqrt{13}$$\times$$\sqrt{13}$ CDW unitcell.
The circle size in {\bfseries\sffamily b} and {\bfseries\sffamily e} is proportional to the amount charge localized in the specific Ta atoms which are marked by the red circle and ellipse in {\bfseries\sffamily a} and {\bfseries\sffamily d}, respectively.
The red and blue circles in {\bfseries\sffamily d} and {\bfseries\sffamily f} denote the majority and minority spin states, respectively.
}
\end{figure} 

Figure 1a shows the prevailing cation(Ta)-centered model of 1T-TaSe$_{2}$, which is the same as the David-star cluster model of 1T-TaS$_{2}$.
One unpaired 5$d$ electron per the CCDW unitcell is localized at the center of a David star and a nondispersive partially filled band appears at the Fermi level (Fig. 1b).
In clear contrast to the case of monolayer 1T-TaS$_{2}$, the partially filled band deviates largely from the center of the band gap originating from the CDW distortion. 
This band overlaps with the valence band edge coming from Se 3$p$ orbitals, which makes a huge difference of the electronic behavior of 1T-TaSe$_{2}$ and 1T-TaS$_{2}$ in their bulk forms. 
In Fig. 1c, the half filled bands splits into upper and lower Hubbard states by onsite Coulomb repulsion (U = 2.3 eV). 
The upper Hubbard state locates at 0.1 eV inside the CDW band gap but the lower Hubbard state is strongly hybridized with the valence bands around 0.4 eV.
The band structure corresponds thus not to a Mott insulator but to a charge transfer insulator. 
These two states are strongly spin-polarized and the ground state predicted in DFT+U is a ferromagnetic or an antiferromagnetic state. 
That is, the band gap itself has a exchange splitting origin in large parts. 

\begin{figure}[t]
\centering{ \includegraphics[width=8.5 cm]{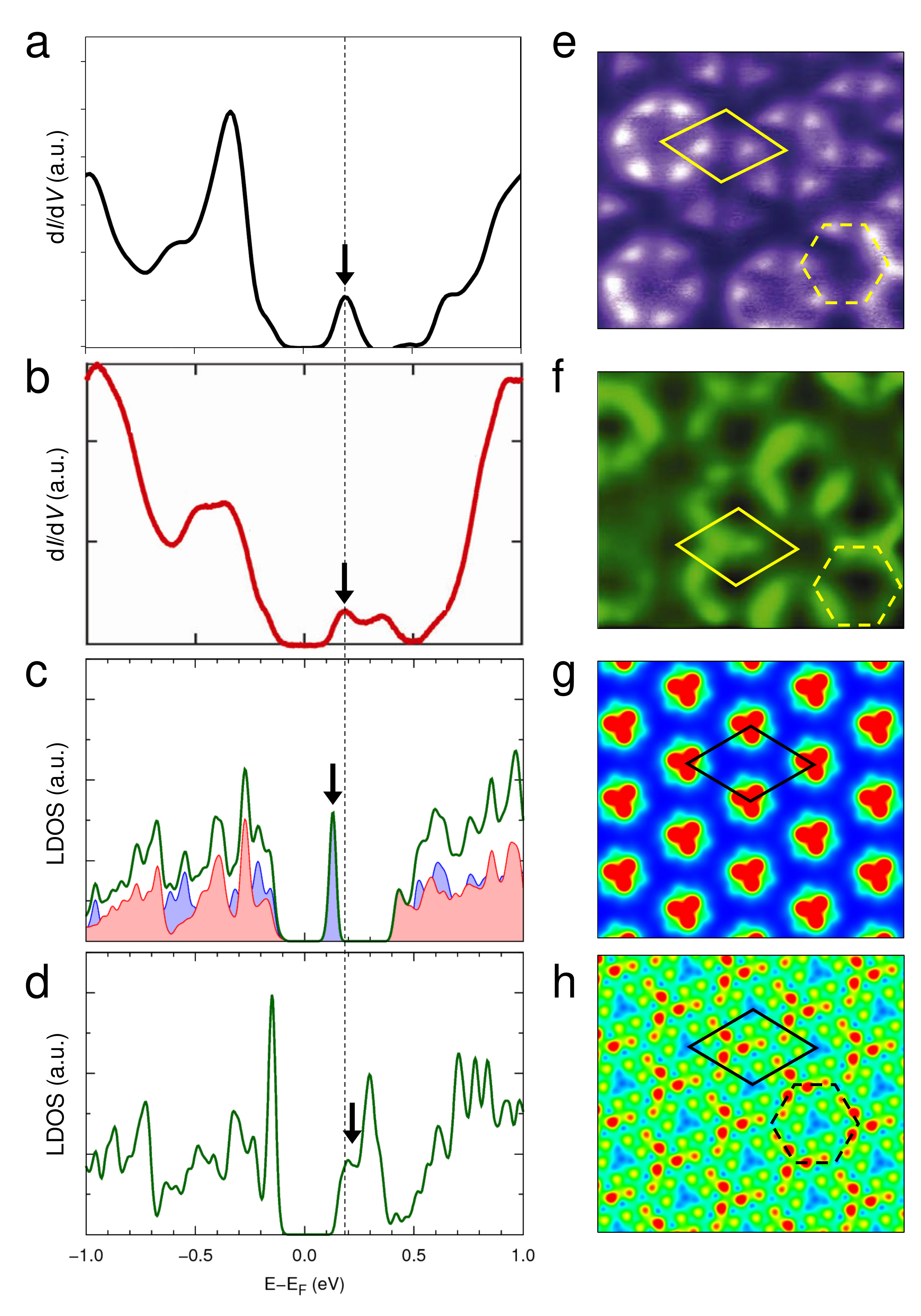} }
\caption{ \label{fig2}
Experimental and theoretical orbital textures for the monolayer 1T-TaSe$_{2}$.
{\bfseries\sffamily a} and {\bfseries\sffamily b} STS spectra of the monolayer 1T-TaSe$_{2}$ on bilayer graphene/SiC(0001) substrate.
{\bfseries\sffamily a} and {\bfseries\sffamily b} were taken from Ref. \cite{chen20} and \cite{lin20}, respectively.
{\bfseries\sffamily c} Local DOS at Ta atoms of cation centered model with electron correlation.
The red and blue shades denote the majority and minority spin states, respectively.
{\bfseries\sffamily d} Local DOS at Ta atoms of anion centered model without electron correlation. The Fermi level is shifted up by 0.15 eV.
{\bfseries\sffamily e-f} d$I$/d$V$ maps and {\bfseries\sffamily g-h} theoretical charge characters.
The arrows in {\bfseries\sffamily a-d} denote the energy level of the corresponding d$I$/d$V$ maps and the theoretical charge characters.
}
\end{figure} 

\begin{figure*}[t]
\centering{ \includegraphics[width=16 cm]{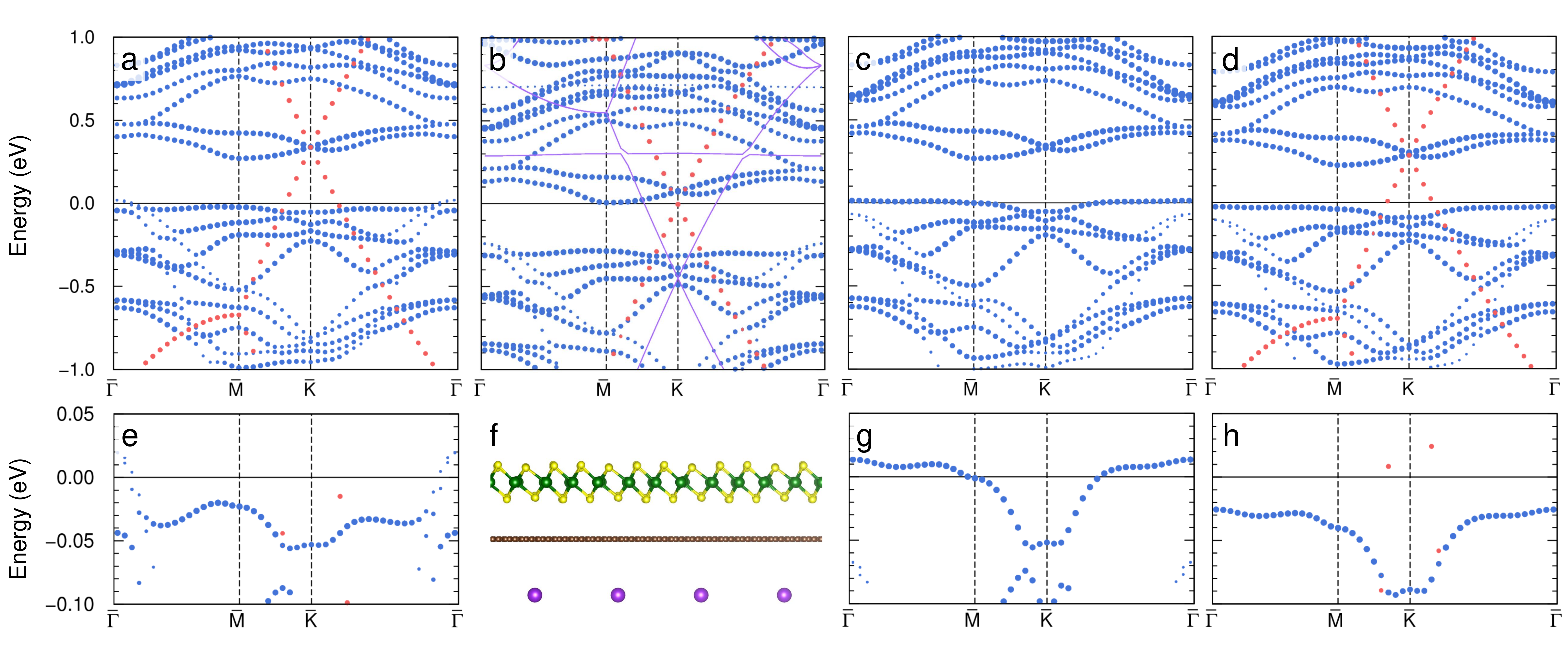} }
\caption{ \label{fig3}
Band structures of the anion centered 1T-TaSe$_{2}$-$\sqrt{13}$$\times$$\sqrt{13}$ on the graphene-5$\times$5.
{\bfseries\sffamily a} 1T-TaSe$_{2}$/graphene,
{\bfseries\sffamily b} 1T-TaSe$_{2}$/graphene/K,  
{\bfseries\sffamily c} 1T-TaSe$_{2}$ (a$_{0}$ = 3.47 \AA), and
{\bfseries\sffamily d} 1T-TaSe$_{2}$/graphene (a$_{0}$ = 3.47 \AA).
The blue and red circles denote the localized states at the Ta and graphene layer, respectively.
The purple solid lines in {\bfseries\sffamily b} denote the band structure of the graphene/K system. 
{\bfseries\sffamily e, g, h} show in enlarged scale near the Fermi level of {\bfseries\sffamily a, c, d}.
{\bfseries\sffamily f} Atomic structure of 1T-TaSe$_{2}$/graphene/K. 
Brown and purple balls represent the C and K atoms, respectively.
}
\end{figure*} 

For the 1T phase, the anion (Se) site is another unique symmetric site of a CDW cluster center to result in a different CDW structure \cite{gye19}.
After the full relaxation with a center at an anion site, 12 Ta atoms are clustered to form a 3-folded irregular hexagon and one Ta atom is left over in between them (Fig. 1d).
The band structure of the anion centered model (Fig. 1e) also exhibits a partially filled metallic state due to thirteen electrons per a $\sqrt{13}$$\times$$\sqrt{13}$ CDW unitcell.
The partially filled state, however, is not a localized state at a single Ta atom but originates from the bonding state of the outer Ta atoms of the cluster marked by red ellipse in Fig. 1d.
A spin polarized ground state is predicted by adding the onsite Coulomb repulsion (Fig. 1f), but the energy levels are changed only marginally.
This indicates that the electron correlation effect is not as crucial as the cation centered model even though the partially filled one-electron band has a very similar band width in both structures. 
That is, the Mott insulator nature is not simply decided by the U/t (t, band width) ratio but the detailed local orbital characteristics has to be considered. 

The discrepancy between the calculation and the experiment mentioned above for the cation-centered model is easily explained in the anion-centered model. 
The corresponding LDOS calculation as shown in Fig. 2d reproduces well the overall STS spectral features (see Supplemental Material Fig. 1 \cite{supp}).
Moreover, the honeycomb orbital characteristic of the state at 0.2 eV is well reproduced (Fig. 2h). 
Two protrusions in the unit cell observed for this state correspond to the top Se atoms at two distinguished regions between the CDW clusters marked by purple shade in Fig. 1d and it explains the double peaks in Fig. 2d.
Thus we suggest that the unusual orbital texture of this system may not be due to an unknown manybody physics but could be explained by a different CDW structure, which results in a simple band insulator.
This orbital character does not change much after inclusion of the correlation U (Fig. 1f).

Since this calculation without U predicts a metallic system as shown in Fig. 1e, we shift the Fermi level by 0.15 eV in the above comparison.
That is, we assume electron transfer from the highly metallic substrate of bilayer graphene on SiC. 
This is plausible since the work function of TaSe$_{2}$ ($\sim$5.1 eV) \cite{tsai19} is larger than that of bilayer graphene/SiC(0001) ($\sim$ 4.3 eV) \cite{mamm17}.
We examined the doping effect for a monolayer 1T-TaSe$_{2}$ in a $\sqrt{13}$$\times$$\sqrt{13}$ structure on the graphene as the most simplest model.
The 5$\times$5 supercell of graphene is matched with the $\sqrt{13}$$\times$$\sqrt{13}$ cell of 1T-TaSe$_{2}$.
In this model, the lattice constant of graphene in the supercell is 2.48 {\AA} which is marginally (less than 1 $\%$) expanded compared to an ideal graphene layer and the interlayer spacing between Ta and graphene layers is optimized to 6.1 {\AA} with an adsorption energy of 0.268 eV per a $\sqrt{13}$$\times$$\sqrt{13}$ unitcell (see Supplemental Material Figs. 2 and 4 \cite{supp}).
While the present calculations do not include the Van der Waals energy for simplicity, the electronic structure is shown to be not sensitive to the interlayer spacing (see Supplemental Material Fig. 3 \cite{supp}). 
The interaction between the 1T-TaSe$_{2}$ and the graphene layer induces a rigid shift upward of the graphene $\pi$ bands by 0.34 eV and most of the partially filled states of 1T-TaSe$_{2}$ are shifted down under the Fermi level (Fig. 3a).
This unambiguously indicates that electrons moves from graphene to 1T-TaSe$_{2}$ without any strong hybridization.
The remaining metallic state of the valence band edge near the $\Gamma$ point is found to be very sensitive to the variation in the lattice constant.
The system can easily be driven to a complete insulating state by only 1 $\%$ lattice expansion (Fig. 3c and d).

While the above discussion deals with a neutral graphene layer, we also examine an electron-doped graphene layer with the Fermi level shifted by -0.45 eV.
This corresponds to the graphene layer grown in SiC(0001) as used to grow 1T-TaSe$_{2}$ layers in the experiment. 
The strongly doped layer was mimicked by putting akali metal (K) adsorbates on the bottom of graphene layer (Fig. 3f) at a distance of 6 {\AA} away (see Supplemental Material Fig. 5 \cite{supp}).  
In this calculation, the doping effect is enhanced, the Dirac point of graphene is shift upward by 0.41 eV (Fig. 3b) and valence band edge of Ta locates at around -0.3 eV.
Thus, we can conclude that the electron-rich graphene layer on SiC(0001) acts as a good electron donor and the partially filled states of the 1T-TaSe$_{2}$ would eliminated by electron doping.

The anion centered model is slightly unfavored in energy compared to the cation centered model as 0.023 eV/Ta atom (0.017 eV/Ta atom) without (with) electron correlation.
Note that the growth condition in molecular beam epitaxy is far from equilibrium and the substrate effect, strain, Moir{\'e} distortion, and charge transfer, can rather easily change the energetics. 
We thus argue that the claim of an unprecedented orbital physics of a Mott insulator in monolayer 1T-TaS$_{2}$ is premature and the careful investigation of the atomic structure and the electron doping from the substrate are requested.
This work indicates that a subtle balance between various different degrees of freedom is needed to secure a Mott insulating state in a complex material. 

This work was supported by Institute for Basic Science (Grant No. IBS-R014-D1). 

\newcommand{\PRL}[3]{Phys.\ Rev.\ Lett.\ {\bf #1}, #2 (#3)}
\newcommand{\PRB}[3]{Phys.\ Rev.\ B\ {\bf #1}, #2 (#3)}
\newcommand{\NA}[3]{Nature\ {\bf #1}, #2 (#3)}
\newcommand{\NAP}[3]{Nat.\ Phys.\ {\bf #1}, #2 (#3)}
\newcommand{\NAM}[3]{Nat.\ Mater.\ {\bf #1}, #2 (#3)}
\newcommand{\PBC}[3]{Physica\ B+C\ {\bf #1}, #2 (#3)}
\newcommand{\RPP}[3]{Rep.\ Prog.\ Phys. \ {\bf #1}, #2 (#3)}
\newcommand{\RMP}[3]{Rev.\ Mod.\ Phys. \ {\bf #1}, #2 (#3)}


\begin{thebibliography}{}
\bibitem{mott68} N. F. Mott, Metal-insulator transition, \RMP{40}{677}{1968}.
\bibitem{cao18} Y. Cao, V. Fatemi, S. Fang, K. Watanabe, T. Taniguchi, E. Kaxiras, and P. Jarillo-Herrero, Unconventional superconductivity in magic-angle graphene superlattices, \NA{556}{43}{2018}.
\bibitem{brou80} R. Brouwer and F. Jellinek, The low-temperature superstructures of 1T-TaSe$_2$ and 2H-TaSe$_2$, \PBC{99}{51}{1980}.
\bibitem{faze79} P. Tosatti, and E. Electrical structural and magnetic properties of pure and doped 1T-TaS$_{2}$. Philos. Mag. B  {\bf 39}, 229 (1979).
\bibitem{salv77} F. J. Di Salvo and J. E.  Graebner, The low temperature electrical properties of 1T-TaS$_{2}$. Solid State Commun. {\bf 23}, 825 (1977).
\bibitem{ge10} Y. Ge and Amy Y. Liu, First-principles investigation of the chargedensity-wave instability in 1T-TaSe$_{2}$, \PRB{82}{155133}{2010}.
\bibitem{chen20} Y. Chen $et$ $al.$, Strong correlations and orbital texture in single-layer 1T-TaSe$_{2}$ \NAP{16}{218}{2020}.
\bibitem{lin20} H. Lin $et$ $al.$, Scanning tunneling spectroscopic study of monolayer 1T-TaS$_{2}$ and 1T-TaSe$_{2}$, Nano Res. {\bf 13}, 133 (2020).
\bibitem{walk81} M. B. Walker and A. E. Jacobs, Distinct commensurate charge-density-wave phases in the 2H-TaSe$_{2}$, \PRB{24}{6670}{1981}.
\bibitem{pore14} M. Porer $et$ $al.$, Non-thermal separation of electronic and structural orders in a persisting charge density wave, \NAM{13}{857}{2014}.
\bibitem{zhen18} F. Zheng, Z. Zhou, X. Liu, and J. Feng, First-principles study of charge and magnetic ordering in monolayer NbSe$_{2}$, \PRB{97}{081101(R)}{2018}.
\bibitem{gye19} G. Gye, E. Oh, and H. W. Yeom, Topological landscape of competing charge density waves in 2H-NbSe$_{2}$,  \PRL{122}{016403}{2019}.
\bibitem{kres96} G. Kresse and J. Furthm{\"u}ller, Efficient iterative schemes for $ab$ $initio$ total-energy calculations using a plane-wave basis set, \PRB{54}{11169}{1996}.
\bibitem{perd96} J. P. Perdew, K. Burke, and M. Ernzerhof, Gerneralized gradient approximation made simple, \PRL{77}{3865}{1996}.
\bibitem{bloc94} P. E. Blochl, Projector augmented-wave method, \PRB{50}{17953}{1994}.
\bibitem{tsai19} H.-S. Tsai $et$ $al.$, Direct synthesis of large-scale multilayer TaSe$_{2}$ on SiO$_{2}$/Si using ion beam techonology, ACS Omega {\bf 4}, 17536 (2019).
\bibitem{mamm17} S. Mammadov, $et$ $al.$ Work function of graphene multilayers on SiC(0001), 2D Mater. {\bf 4}, 015043, (2017).
\bibitem{supp} Supplemental Material

\end{thebibliography}
\end{document}